\documentclass[useAMS,usenatbib]{mn2e}
\usepackage{graphicx,amssymb,epsfig}
\usepackage{aas_macros}
\usepackage{natbib}
\newcommand{\kms}{\,{\rm km\,s^{-1}}}
\newcommand{\msun}{\,{\rm M_\odot}}

\newcommand{\vir}{{\rm vir}}

\newcommand{\beq}{\begin{equation}}
\newcommand{\eeq}{\end{equation}}
\newcommand{\ba}{\begin{eqnarray}}
\newcommand{\ea}{\end{eqnarray}}
\def\spose#1{\hbox to 0pt{#1\hss}}
\newcommand{\lta}{\mathrel{\spose{\lower 3pt\hbox{$\mathchar"218$}}
      \raise 2.0pt\hbox{$\mathchar"13C$}}}
\newcommand{\gta}{\mathrel{\spose{\lower 3pt\hbox{$\mathchar"218$}}
      \raise 2.0pt\hbox{$\mathchar"13E$}}}
\def\simlt{\mathrel{\rlap{\lower 3pt\hbox{$\sim$}}\raise 2.0pt\hbox{$<$}}}
\def\simgt{\mathrel{\rlap{\lower 3pt\hbox{$\sim$}} \raise 2.0pt\hbox{$>$}}}

\title[Black holes in dwarfs]
{Massive black holes lurking in Milky Way satellites}

\author[Van Wassenhove et al]{S. Van Wassenhove$^{1}$, M.Volonteri$^{1}$, M. G. Walker$^{2}$  \& J. R. Gair$^{2}$\\ 
$^{1}$Department of Astronomy, University of Michigan, Ann Arbor, MI, USA\\
$^{2}$Institute of Astronomy, Madingley Road, Cambridge CB3 0HA, UK\\}
\begin{document}
\maketitle
\begin{abstract}
As massive black holes (MBHs) grow from lower-mass seeds, it is natural to expect that a leftover population of progenitor MBHs should also exist in the  present universe.   Dwarf galaxies undergo a quiet merger history, and as a result, we expect that dwarfs observed in the local Universe retain some `memory' of the original seed mass distribution.  Consequently, the properties of MBHs in nearby dwarf galaxies may provide clean indicators of the efficiency of MBH formation.  In order to examine the properties of MBHs in dwarf galaxies, we evolve different MBH populations within a Milky Way halo from high-redshift to today. We consider two plausible MBH formation mechanisms: `massive seeds' formed via gas-dynamical instabilities and a Population III remnant seed model. `Massive seeds' have larger masses than PopIII remnants, but form in rarer hosts. We dynamically evolve all halos merging with the central system, taking into consideration how the interaction modifies the satellites, stripping their outer mass layers. We compare the population of satellites to the results of N-body simulations and to the observed population of dwarf galaxies. We find good agreement for the velocity, radius and luminosity distributions.  We compute different properties of the MBH population hosted in these satellites. We find that some MBHs have been completely stripped of their surrounding dark matter halo, leaving them `naked.'   We find that for the most part MBHs retain the original mass, thus providing a clear indication of what the properties of the seeds were. We derive the black hole occupation fraction (BHOF) of the satellite population at $z=0$. MBHs generated as `massive seeds' have large masses that would favour their identification, but their typical BHOF is always below 40 per cent and decreases to $\la 1$ per cent for observed dwarf galaxy sizes.  In contrast, Population III remnants have a higher BHOF, but their masses have not grown much since formation, inhibiting their detection. 
\end{abstract}

\section{Introduction}
Deriving observational constraints on MBH formation, occurring at very high redshifts, is clearly challenging. The first `seeds' could be as light as a few  hundred $\msun$.  Also, initial conditions  tend to be erased very fast if accretion is efficient in growing MBHs \citep{VG2009}. {\it Galactic archeology} offers an alternative to high-redshift observations.  Simple arguments suggest that MBHs might inhabit also the nuclei of dwarf galaxies, such as the satellites of the Milky Way and Andromeda, today. As MBHs grow from lower-mass seeds, it is natural to expect that a leftover population of progenitor MBHs should also exist in the present universe.   Indeed, we expect that one of the best diagnostics of `seed' formation mechanisms would be to measure the masses of MBHs in dwarf galaxies.  The progenitors of massive galaxies have in fact a high probability that their central MBH is not `pristine', that is, it has increased its mass by accretion, or it has experienced mergers and dynamical interactions. Any dependence of the MBH mass, $M_{\rm BH}$, on the initial seed mass is largely erased.  In contrast, dwarf galaxies undergo a quieter merger history, and as a result, at low masses the  MBH occupation fraction and the distribution of MBH masses still are expected to retain some `memory' of the original seed mass distribution. The signature of the efficiency of the formation of MBH seeds will consequently be stronger in dwarf galaxies \citep{VLN2008}.  

The records for the smallest known MBHs belong to the dwarf Seyfert~1 galaxies POX 52 and NGC 4395. They are believed to contain black holes of mass $M_{BH} \sim 10^5\,M_\odot$ \citep{barthetal2004, Peterson2005}.   There are also significant non-detections of MBHs in a few nearby galaxies from stellar-dynamical observations, most notably the Local Group Scd-type spiral galaxy M33, in which the upper limit to $M{\rm BH}$ is just a few thousand solar masses \citep{Gebhardt2001,Merritt2001}.  Similarly, in the Local Group dwarf elliptical galaxy NGC 205, $M_{\rm BH} < 3.8\times10^4 \msun$ \citep{Valluri2005}.  These results suggest that the MBH `occupation fraction' in low-mass galaxies might be significantly below unity, but at present it is not possible to carry out measurements of similar sensitivity for galaxies much beyond the limits of the Local Group \citep{Ibata2009,Lora2009,Greene2007,Debattista2006}. 

In this paper we explore theoretical expectations for (1) the probability that a dwarf galaxy hosts an MBH, and (2) if the merging and accretion history in dwarf galaxies leads to different scaling relationships of MBHs with their hosts. We explore here the predictions of different models of MBH `seed' formation, and derive constraints that may be testable with current and future instruments. 

\section{Massive black hole formation and dynamical evolution}

We follow the formation and evolution of an MBH population in a Milky Way-size halo in a $\Lambda$CDM universe \citep{WMAP5}. Our technique follows that of \cite{VHM}, as we use Monte Carlo realisations of the merger histories of dark matter halos. We analyse here 5 different realisations  of halos that reach a mass of $M_h=2 \times10^{12}\msun$ at $z=0$. We seed the  high-redshift progenitor halos with black holes and follow them from formation to $z=0$. We focus on the signatures of black hole formation efficiency in satellite galaxies surviving until today.

\subsection{Massive black hole formation models}
We adopt three different models to seed haloes with black holes: massive seeds and two Population III remnant seed models. These models determine which haloes are seeded with black holes at high redshifts. They also set a minimum mass for central MBHs today. 

\subsubsection{Population III remnants}
For the Population III remnant models, we follow two schemes that differ only in the efficiency of MBH seed formation. We assume here that MBHs form as end-product of the very first  generation of stars. The main features of a scenario for the hierarchical assembly of MBHs left over by the first stars in a $\Lambda$CDM cosmology have been discussed by \cite{VHM}, \cite{VRees2006} and \cite{VN09}.  The first stars are believed to form at $z\sim 20-30$ in minihalos, $M_{\rm min}\approx  10^6\,\msun$, above the cosmological  Jeans mass. They collapse at $z\sim 20-50$ from the rarest  $\nu$-$\sigma$ peaks of the primordial density field.  In this regime,  cooling is possible by means of molecular hydrogen \citep[$T_{\rm vir} > 2-3 \times10^3$ K,][]{Tegmark1997,Yoshida2006}, but the inefficient cooling at zero metallicity might lead to a very  top-heavy initial stellar mass function.  Specifically, the earliest-forming stars are likely to have been very massive \citep{CBA84,bromm1999,bromm2001,abel2000,Yoshida2006}. If stars form above 260 $M_\odot$,  they  would rapidly collapse to massive black holes with little mass loss \citep{fryer2001},  i.e., leaving behind seed MBHs with masses $M_{BH} \sim 10^2-10^3\,M_\odot$  (Madau \& Rees 2001).  We here consider that MBH seeds populate haloes with formation redshift $z>12$ which represent density peaks with $\nu_c=3$  (`peak3'; as a reference: $M_{\rm h}>10^5 \msun$ at $z=20$, and $M_{\rm h}>10^8\msun$ at $z=12$) or $\nu_c=3.5$ (`peak3.5';  as a reference: $M_{\rm h}>10^6 \msun$ at $z=20$, and $M_{\rm h}>10^9\msun$ at $z=12$), while also  requiring that $T_{\rm vir}\simgt 2500$ K to ensure effective H$_2$ cooling (implying masses above $3\times 10^6 \msun$).   
Seeds form as 100$\msun$ black holes. 

\subsubsection{Massive MBH seeds}
The massive seed model relies instead on the collapse of supermassive objects formed directly out of dense gas \citep[and references therein]{Koushiappas2004, BVR2006,LN2006}, where the mass inflow is regulated by the degree of stability of the cooling gas. Here we assume that gas is accumulated in the centre of a halo via viscous instabilities.   A dynamically unstable disc can develop non-axisymmetric spiral structures that effectively redistribute angular momentum, causing mass inflow.   This process stops when the amount of mass transported to the centre is enough to make the disc marginally stable. We here follow \cite{LN2006}, who suggest that the mass inflow can be computed from the Toomre stability criterion and from the disc properties, determined from the dark matter halo properties \citep[halo mass, $M_{\rm h}$, virial temperature, $T_{\rm vir}$ and spin parameter, $\lambda$;][]{MoMaoWhite1998}. We refer the reader to \cite{LN2006,VLN2008} for a comprehensive description.  We summarize here the main features of the model, and how we operationally implement it. 

Consider a dark matter halo of mass $M_{\rm h}$ and virial temperature $T_{\rm vir}$\footnote{A halo at redshift $z$ is uniquely characterized by a virial radius  $r_\vir$, defined as the radius of the sphere encompassing a mean mass overdensity $\delta_\vir$. $\delta_{\rm vir} \approx 178$ in an Einstein-de Sitter Universe.  Detailed calculations for different cosmologies (e.g., $\Lambda$CDM) can be found in \cite{Lacey1993} and \cite{Eke1996}. From the virial theorem the virial mass, $M_{\rm h}$ can be calculated straightforwardly, along with the circular velocity, $V_c=\sqrt{GM_{\rm h}/r_\vir}$ and virial temperature $T_\vir=\mu\, m_p V^2_c/(2 k_B)$ , where $\mu\simeq 0.722$ is the mean molecular weight, $m_p$ is the proton mass and $k_B$ is the Boltzmann constant. }, containing a gas mass in cold gas $M_{\rm gas}=f_{\rm d}M_h$ (we assume that the gas fraction cooling in is roughly 5\% implying $f_{\rm d}=0.05$, \citealt{MoMaoWhite1998}).  The other main parameter characterizing a dark matter halo that is relevant here is its spin parameter $\lambda$ ($\equiv J_h E_h^{1/2}/ GM_h^{5/2}$, where $J_h$ is the total angular momentum and $E_h$ is the binding energy).  The distribution of spin parameters for dark matter halos  measured in numerical simulations is well fit by a lognormal distribution in $\lambda_{\rm spin}$, with mean $\bar \lambda_{\rm spin}=0.05$ and standard deviation $\sigma_\lambda=0.5$  (e.g., \citealt{Bullock2001, VandenBosch2002, Maccio2008}).

If the virial temperature of the halo $T_{\rm vir}>T_{\rm gas}$, the gas collapses and forms a rotationally supported disc.  For metal--free gas, the cooling function is dominated by hydrogen. In thermal  equilibrium, if the formation of molecular hydrogen is suppressed (see the discussion in \citealt{Devecchi2009}),  these discs are expected to be nearly isothermal at a temperature of a few thousand Kelvin (here we take $T_{\rm gas}\approx 5000$K, \citealt{LN2006}).   For low values of the spin parameter $\lambda$, the resulting disc can be compact and dense and is subject to gravitational instabilities. This occurs when the Toomre stability parameter $Q$ approaches a critical value $Q_{c}$ of order unity (following Volonteri et al. 2008 we adopt $Q_c=2$, in order to match observational constraints on the MBH and quasar population). If the destabilization of the system is not too violent, instabilities lead to mass infall instead of fragmentation into bound clumps and global star formation in the entire disc \citep{LN2006}.    This is the case if the inflow rate is below a critical threshold $\dot{M}_{ max}=2\alpha_{c}\frac{c^3_s}{G}$ that the disk is able to sustain \citep[where $\alpha_c\,\sim\, 0.12$ describes the viscosity,][]{RLA05} and molecular and metal cooling are not important. 

This mass redistribution process ceases when the amount of mass transported to the center, $M_a$, is enough to make the disc marginally stable. This can be computed easily from the Toomre stability criterion and disc properties, determined from the dark matter halo mass
and angular momentum \citep{MoMaoWhite1998}:
\begin{equation}
M_{a}= \left\{\begin{array}{ll}
\displaystyle m_{\rm d}M_h\left[1-\sqrt{\frac{8\lambda}{m_{\rm d}Q_{\rm c}}
\left(\frac{j_{\rm d}}{f_{\rm d}}\right)\left(\frac{T_{\rm gas}}{T_{\rm
vir}}\right)^{1/2}}\right] & \lambda<\lambda_{\rm max} \\
0 & \lambda>\lambda_{\rm max}
\nonumber
\end{array}
\right.
\label{mbh}
\end{equation}
where 
\begin{equation}
\lambda_{\rm max}=f_{\rm d}Q_{\rm c}/8(f_{\rm d}/j_{\rm d}) (T_{\rm
  vir}/T_{\rm gas})^{1/2}
\label{lambdamax} 
\end{equation}
is the maximum halo spin parameter for which the disc is gravitationally unstable, and $j_d$ is the fraction of the halo angular momentum retained by the collapsing gas ($j_d =f_d$ if specific angular momentum is conserved). 

For large halo masses, the internal torques needed to redistribute the excess baryonic mass become too large to be sustained by the disc, which then undergoes fragmentation. This occurs when the virial temperature exceeds a critical value $T_{\rm max}$, given by:
\begin{equation}
\frac{T_{\rm max}}{T_{\rm gas}}=\left(\frac{4\alpha_{\rm c}}{f_{\rm
d}}\frac{1}{1+M_{\rm BH}/f_{\rm d}M_h}\right)^{2/3}.
\label{frag}
\end{equation}

To summarize, every dark matter halo is characterized by its mass $M_h$ (or virial temperature $T_{\rm vir}$) and by its spin parameter $\lambda$. The latter is drawn from a lognormal distribution in $\lambda_{\rm spin}$ with mean $\bar \lambda_{\rm spin}=0.05$ and standard deviation $\sigma_\lambda=0.5$ \citep[][and references therein]{Maccio2007}. The gas has a temperature $T_{\rm gas}=5000$K. If $\lambda<\lambda_{\rm max}$ (see Eq~\ref{lambdamax}) and $T_{\rm vir}<T_{\rm max}$ (Eq.~\ref{frag}), then we assume that a seed BH of mass $M_{\rm BH}=M_a$ given by eqn.~(\ref{mbh}) forms in the center. 

The gas made available in the central compact region can then form a central massive object, for instance via the intermediate stage of a `supermassive' star \citep{hoyle1963,Baumgarte1999}, or a `quasistar' \citep[an initially low-mass black hole rapidly accreting within a massive, radiation-pressure-supported envelope, see also][]{BVR2006,Begelman2008}. Hence, the black hole seed mass estimates based on Equation~\ref{mbh} should be considered  as upper limits. We here consider that MBH seeds with $M_{\rm BH}\simeq M_{\rm a}$ can form in haloes with formation redshift $z>12$ that satisfy all the above criteria. The mass function of seeds peaks at $10^5-10^6 \sun$ \citep[see Figure~2 in][]{VLN2008}.

We remind the reader here that this process is effective for halos with low angular momentum (low spin parameter, $\lambda \simlt 0.01$) and zero metallicity, where cooling is driven by atomic hydrogen cooling and the difference between gas and virial temperature is small (making the disc resilient to global fragmentation and star formation). 
The efficiency of the seed assembly process ceases at large halo masses ($T_{\rm vir}>1.4\times10^4$ K), where the mass-accretion rate from the halo is above the critical threshold for fragmentation and the disc undergoes global star formation instead.

\subsection{Massive black hole and galaxy evolution}
To study the MBHs at $z=0$, we follow the evolution of the black holes along with their host galaxies using models that track both black hole mergers and accretion.  Several recent works point out how fragile the environment of low-mass galaxies is \cite[and references therein]{Bovill2009}.  The shallow potential well of these galaxies makes it easy for gas to evaporate or escape in the presence of feedback or dynamical heating.  We incorporate a simple scheme that tracks the gaseous content of  galaxies and the effect of gas depletion on MBH growth and dynamical evolution. 

\subsubsection{Halo baryon content}
We track the baryon content of haloes using a method similar to that developed by \cite{Okamoto09}.  They present a simple model that reproduces the results of cosmological hydrodynamical simulations of mass loss from a UV background.  In this model, haloes form at high redshifts with the cosmic mean baryon fraction, $f_b=0.18$ \citep{WMAP5}.  We define the baryon fraction as the ratio between the mass in baryons and the total mass of a halo.  After reionization (set at $z=9$ here), low mass haloes lose their baryons as a result of the increased temperature of the intergalactic medium (IGM) \citep{gnedin2000}.  The condition for baryon loss depends on the relation of the virial temperature of the halo, $T_{\rm vir}$, to the temperature at which photoheating and radiative cooling are balanced, $T_{\rm eq}$.  We evaluated $T_{\rm eq}$ at an overdensity, $\Delta_{\rm evp} = 10^6$, representing the densest, most bound region of the halo \citep[see][for a discussion of the model parameters and the allowed range]{Okamoto09}.  If the equilibrium gas temperature in these dense regions is greater than the virial temperature of the halo, the gas will evaporate out of the halo into the surrounding IGM.  In more massive haloes, the dense gas is able to cool efficiently, preventing it from evaporating.  We model $T_{\rm eq}$ using the UV background by \cite{HaardtMadau96}. Halos with $T_{\rm vir} < T_{\rm eq}(\Delta_{\rm evp})$ have a mass in baryons, $M_{\rm b}$, which decreases with time, given by:
\begin{equation}
M_{\rm b}(t + \delta t) = M_{\rm b}(t)e^{-\frac{\delta t}{t_{\rm evp}}}.
\end{equation}
The evaporation time-scale, $t_{\rm evp}$, is given by $R_{\rm vir}/c_{\rm s}(\Delta_{\rm evp})$, where $R_{\rm vir}$ is the virial radius of the halo, and $c_{\rm s}(\Delta_{\rm evp})$ is the sound speed at the evaporation overdensity.  This timescale corresponds to the time for gas to leave the halo moving at the sound speed.

We additionally allow for accretion of baryons from the IGM onto haloes.  If gas at the outskirts of the halo is colder than the virial temperature of the halo, the halo accretes enough baryons to reach the cosmic mean.  Here the temperature of the accreting gas is approximated as the equilibrium temperature of gas with density $\rho_{\rm vir}/3$.  This density corresponds to the density of gas at the virial radius of the halo.  The requirement for accretion up to the cosmic mean baryon fraction is expressed $T_{\rm eq}(\rho_{\rm vir}/3) < T_{\rm vir}$.  When the halo is colder than the gas in the surrounding IGM, no accretion occurs.  When two haloes merge in our model, the resulting halo has a mass in baryons equal to the sum of the baryonic masses of the progenitor haloes. 

\subsubsection{Massive black hole accretion}
Based on  simulations of MBH mergers in galaxies with different gaseous content \citep{Callegari2009}, we assume that black holes hosted by a baryon rich halo experiencing a major merger (mass ratio greater than 1:10) accrete mass from the host. We here define a halo as baryon rich when it has retained more than half of its original baryon fraction. Specifically, we choose a baryon fraction threshold $f_b>0.1$.  In this simple treatment, a merged black hole accretes mass according to the central velocity dispersion of the host $M_{bh}=10^8(\sigma/200 \kms)^4\msun$ \citep[e.g.,][]{Tremaineetal2002}. We link the central velocity dispersion to the circular velocity of the halo using the empirical relationship $\log(V_c)=0.74\log(\sigma)+0.8$ \citep{Pizzella05}.  The correlation between central velocity dispersion and halo circular velocity has been studied observationally for samples of galaxies mostly in the range $\sigma \gtrsim 70 \kms$ \citep{Ferrarese2002,Baes2003}. Pizzella et al. (2005) extend the study to $\sigma \sim 40 \kms$, but extending these studied to the range of dwarf galaxies is challenging, as the stars may not reach the radius where the halo has its maximum circular velocity \citep[e.g., see Fig 5 of][]{Penarrubia2008}.   Since there is no accepted relationship between $V_c$ and the central velocity dispersion that extends down the the dwarf galaxy sizes considered here, we use the above relationship to determine MBH masses from the halo velocity dispersion.  For dynamical evolution of the satellites (cfr. Section 2.3.1), we use the definition $\sigma=V_c/\sqrt[]{3}$, as the velocity dispersions quoted in the literature that we need to assess our dynamical modelling are obtained from the `global' velocity sample in each galaxy, not just from some innermost sub-sample. At present the only dwarf spheroidal (dSph) satellite with maximum circular velocity constrained by data (and assuming constant velocity anisotropy) is Fornax, with global velocity dispersion $\sigma \simeq  11 \kms$ and $V_c \sim 18 \kms$ \citep{walker09a}, consistent with the   $\sigma=V_c/\sqrt[]{3}$ scaling.  We note that our overall results are robust to changes in the assumed scaling between $V_c$ and $\sigma$. We have run models where we also adopt the relationship $\sigma=V_c/\sqrt[]{3}$ to calculate MBH masses, and  qualitatively all results hold.

The total mass of the black hole resulting from a completed merger is the sum of the merging black hole masses and any accreted mass. When a halo with a central black hole merges with an empty halo, the central black hole will accrete mass after the merger timescale if the halo remains baryon rich. A similar scheme has been shown to reproduce observational constraints on MBH evolution \citep[luminosity function of quasars and Soltan's argument, $M_{\rm BH}-\sigma$ relationship at $z=0$, mass density in MBHs at $z=0$;][]{VLN2008} for the `peak3.5' and `massive seeds' models.

 \begin{figure}   
   \includegraphics[width=\columnwidth]{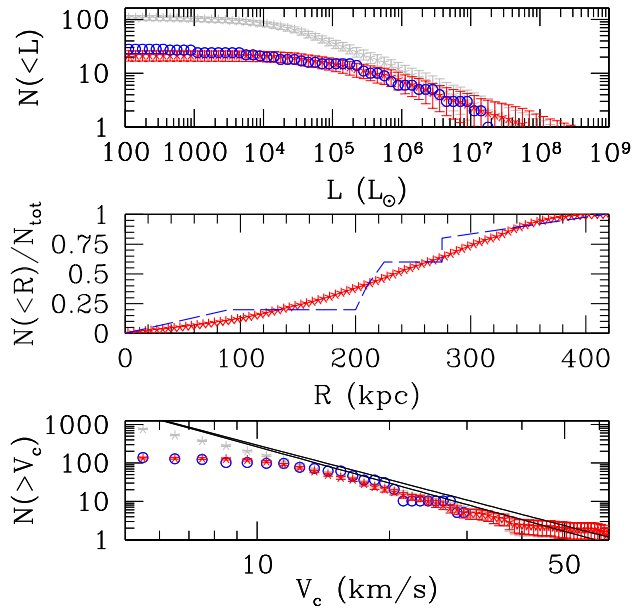} 
   \caption{Top panel: luminosity function of satellite galaxies, corrected to match the SDSS DR5 selection. Upper grey curve with error-bars: full sample of  simulated galaxies surviving tidal stripping. Lower red curve with error-bars: galaxies with baryon fraction $>7.5\times 10^{-3}$. Blue circles: observed luminosity function of Milky Way satellites.  Middle panel: cumulative radial distribution of our simulated satellite population (dashed line: from Tollerud et al. 2008, fig. 5. `DR5' distribution). Bottom panel: cumulative velocity distribution of our simulated satellite population compared to the fit derived for satellites  in Via Lactea I (lower line, Diemand et al.  2007) and Aquarius (upper line, Springel et al. 2008). Circles: observed dwarf spheroidal satellites  of the Milky Way. Errorbars are 1$-\sigma$ Poissonian errors. }
   \label{fig:VL}
\end{figure}

\subsection{Dynamical evolution of satellites and massive black holes}
 As shown by several investigations \citep[e.g.,][]{Madau2008,Tollerud2008,Maccio2010,Munoz2009} the population of  dark matter satellites found in numerical simulations exceeds by a large degree the number of known satellites of the Milky Way, creating a ``missing satellite" problem. It has been advocated that the solution to this problem may lie in a combination of factors. On the one hand cooling and star formation are inefficient in the presence of a strong photoionizing background, which prevents the development of a conspicuous luminous component (cfr. Section 2.2.1). On the other hand,  tidal stripping of the satellites orbiting in the potential of the Milky Way may cause mass loss, leading to systems much less massive than they were at the time that they merged with their host halo. In Section 2.3.1 we describe how we model the dynamical evolution of satellites in the Milky Way halo in order to derive the properties of the sub-population of luminous satellites that have survived until today. In Section 2.3.2 we address the dynamical evolution of MBH pairs formed during galaxy mergers.

\subsubsection{Dynamical evolution of satellites}
After a halo merger, the smaller halo becomes a satellite of the more massive system. These satellite haloes evolve in the potential well of the host until $z=0$, experiencing tidal stripping and possibly merging with the host. 

 We model the dynamical evolution of satellites within the host halo potential using analytical techniques \citep{Volonterietal08}. For each satellite that merges with the main halo of the merger tree, we evolve the satellite-host system by integrating the equation of motion of the satellite in the gravitational potential of the host \citep[assuming spherical NFW profiles;][]{NFW1997}, including the dynamical friction term:
\begin{equation}
{d^2 {\vec r} \over d t^2} =  -{G M(r) \over r^2}\, {\vec r} - 
{4 \pi G^2 \ln\Lambda\,\rho\, M_{\rm sat} \over v^2} f(x)\,{\vec v}
\label{DF}
\end{equation}
 where $f(x)\equiv [{\rm erf}(x) - (2 x/\sqrt{\pi}) e^{-x^2}]$, $x\equiv v/\sqrt{2}\sigma$, 
and the velocity dispersion $\sigma$ is derived from the Jeans' equation for the composite density
profile, assuming isotropy (e.g. \citealt{Binney1987}). Here $M(r)$ describes the total 
mass of the host within $r$, $\rho(r)$ is the total density profile, and the second 
term represents dynamical friction against the background. We include the MBH keplerian potential if the galaxies host a MBH.  The Coulomb logarithm, $\ln \Lambda$, in equation (\ref{DF}) is taken equal to 2.5  \citep{Taylor2001}.

The mass of the satellite evolves during the integration because of tidal stripping. At every step of the integration we compare the mean density of the satellite to the mean density of the host halo at the location of the satellite. Tidal stripping occurs at the radius within which the mean density of the satellite exceeds the density of the galaxy interior to its orbital radius \citep{Taylor2001}.   We trace the orbital evolution and the tidal stripping of all satellites from the time when the satellite enters the virial radius of the host to $z=0$. Satellites that survive until the present time provide an analogue of the dwarf galaxy population around the Milky Way. 

We compare the circular velocity and radial distribution of our satellite population to that of Via Lactea and Aquarius simulations \citep{Diemand2007,Madau2008,Springel2008} and with constraints derived from the observed populations \citep{Tollerud2008,Koposov2008,Walsh2009}. Figure~\ref{fig:VL} compares our results to the literature. We find good agreement with the circular velocity distribution of satellites in the Aquarius simulation at $z=0$ at velocities above $V_{max}\ga 10 \kms$. 

We further derive the luminosity function of the satellites by assigning a luminosity based on their velocity dispersion. We adopt an empirical correlations derived by \cite{Kormendy2004} for galaxies with velocity dispersion $\sigma > 30 \kms$:
\beq
\log(L_V/L_\odot)=7.80 + 5\log(\sigma/30\kms).   
\label{eq:kormendy}
\eeq
We perform a least square fit in $\log(L_V)$ vs $\log(\sigma)$ for low $\sigma$ systems (using data from \cite{Mateo:1998}, \cite{Simon2007}  and \cite{Walker2009b}) we find a very similar relationship ($\log(L_V/L_\odot)=(7.6\pm0.6)+ (4.3\pm 0.9) \log(\sigma/30\kms)$), albeit with a larger scatter. We therefore adopt the fit in Equation~\ref{eq:kormendy} for the whole $\sigma$ range. To derive the luminosity function we need to set one single parameter, that is the minimum baryon fraction that allows star formation. We find the best fit to the observed luminosity function by considering as `luminous' only those satellites with a baryon fraction $>7.5\times 10^{-3}$ (the acceptable range is $6.5\times 10^{-3}<f_b<8\times 10^{-3}$). To compute the luminosity function, we follow \cite{Madau2008} in correcting our theoretical sample to match the SDSS DR5 sample of satellites.  First, we correct the sample by a factor $f_{\rm DR5}=0.194$ that takes into account the sky coverage. We then apply a joint distance-magnitude cut \citep{Koposov2008, Tollerud2008} ensuring that we count only satellites within the SDSS completeness limit:
\beq
r_{\rm max}=\left(\frac{3}{4\pi f_{\rm DR5}} \right)^{1/3} \, 10^{(-0.6 M_{\rm V} -5.23)/3} \, {\rm Mpc}.
\eeq

We use the central black holes of this population of haloes to provide observational signatures of black hole formation and growth efficiency. We note that stripping sometimes (a few per cent of the cases) leaves some `naked'  MBHs, as the host loses most of its mass. 

\subsubsection{Dynamical evolution of massive black holes}

Along the dynamical evolution of the MBH+galaxy systems, we further must determine the fate of the MBHs they contain (if any).  Work by Callegari et al. (2009) has shown that the efficiency of black hole pairing is a function both of the mass ratio and the baryon fraction of the merging haloes. Black holes in merging disc galaxies form a bound pair when the mass ratio of the merging galaxies is larger than 1:10 and galaxies are gas-rich (i.e., cold gas represents 10 per cent of the disc mass). We therefore define a major merger to be a merger between haloes with mass ratio greater than 1:10. In our models, minor mergers (mass ratio less than 1:10) do not lead to efficient black hole pairing and mergers. When a major merger occurs, the final fate of the MBHs depends on the baryon content of the host. There is insufficient dynamical friction in gas-poor galaxies to lead to efficient MBH pairing. We assume that black hole mergers stall when occurring in a baryon poor halo ($f_b<0.1$).   Unless stalled, we assume that black holes merge within the merger timescale of their host galaxies. For major mergers, this timescale is well approximated by $\tau_{\rm merge}/\tau_{\rm dyn}\simeq0.4(M_{\rm host}/M_{\rm sat})^{1.3}/ \ln(1+M_{\rm host}/M_{\rm sat})$, as shown by \cite{BKetal08},  where $\tau_{\rm dyn}$ is the dynamical time at the virial radius. This timescale represents a lower limit to the MBH-MBH merger time \citep{BBR1980},  although the assumption that MBHs merge within the merger timescale of their hosts is likely for MBH pairs formed after gas rich galaxy mergers \citep{Escala2005,Dotti2007}.

\section{Black hole occupation fraction and mass scaling}

\begin{figure}   
   \includegraphics[width=\columnwidth]{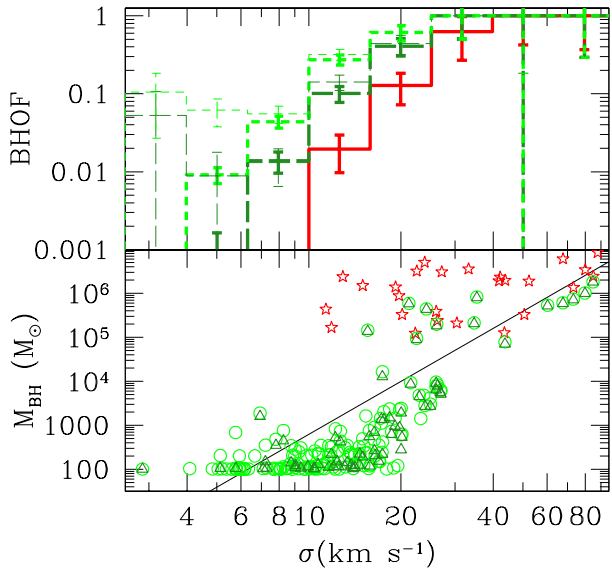} 
   \caption{Top: fraction of galaxies, at a given velocity dispersion, which host a central MBH (occupation fraction, BHOF). Long-dashed green histogram: `peak3.5' case. Short-dashed green histogram: `peak3' case.  Thick lines: all satellites. Thin lines: luminous satellites. 
   Solid red histogram: `massive seeds'.  Bottom: the $M_{\rm BH}-$velocity dispersion ($\sigma$) relationship for MBHs in satellites. We here show the results of a suite of ten realisations. Stars: `massive seeds'. Circles: `peak3.5'. Triangles: `peak3'.}
   \label{fig:BHOF}
\end{figure}

We study the black hole population of satellites at $z=0$ to find signatures of the black hole properties and seed efficiency.  We show the black hole occupation fraction (BHOF) of the satellite population at $z=0$ in Figure~\ref{fig:BHOF}. Massive satellites today were likely to have massive progenitors at high redshifts, meaning that they are more likely to host central black holes, given that a minimum mass threshold is required in all our models (mimicking the necessity of a deep potential well). 
The higher the threshold mass, the more the BHOF is expected to decrease with the mass (or velocity dispersion) of today's host.  As expected, the more biased scenario `peak3.5' leads to a lower overall BHOF than `peak3'. The `massive seed' scheme produces fewer black holes than either `peak3' or `peak3.5'. This is because of the stricter conditions needed for MBH formation: seeds form only in very massive halos  ($M_{\rm h}\simeq 10^8 \msun$, compared to the much lower mass threshold for the Population III remnant case, $M_{\rm h}\sim10^7 \msun$ at $z=15$ for `peak3' and $M_{\rm h}\sim 7\times 10^7 \msun$ for `peak3.5') with low angular momentum (low spin parameter).  Given that spin parameters appear to be distributed lognormally (with mean $\bar \lambda_{\rm spin}=0.05$ and standard deviation $\sigma_\lambda=0.5$) in halos extracted from cosmological simulations, typically  only $\sim$10 per cent of the haloes in the allowed mass range can form a central seed MBH, for the parameter choice discussed in section 2.1. The results shown here are largely insensitive to the efficiency of black hole merging and accretion. The merger and accretion efficiency primarily affect the mass of the central black holes, not the presence or absence of one. The present-day BHOF is therefore a sensitive probe of the efficiency of black hole formation in haloes at high redshifts. Observations of satellites at relatively high velocity dispersions might therefore distinguish between the massive seed and Population III models presented here, although this prospect depends on the ability to detect MBHs where they exist (see Section 4). 

Unlike the BHOF,  the masses of MBHs within galaxies are in general sensitive to the merger efficiency and accretion. However, in the environment we are investigating we expect MBH growth to be largely inefficient.  In our model, black holes accrete mass when the host halo experiences a major merger and remains baryon rich over the merger timescale. Gas-rich major mergers are rare for the progenitors of our satellite galaxies, so MBH growth through mergers and accretion is suppressed.  

In a baryon poor halo, MBH-MBH mergers will not complete, accretion does not occur, and the black hole mass does not change. Even before reionization, when all galaxies are gas-rich, two factors contribute to the limited growth of MBHs. On the one hand,  the rarity of seeds causes MBHs to evolve mostly in isolation.  On the other hand, MBH growth through accretion at high redshifts is negligible in our model. Accreting MBHs grow according to the velocity dispersion of the host halo, but the seed masses we consider exceed the MBH mass expected for small haloes.  For satellites with velocity dispersions similar to those of the Milky Way dwarf galaxies, Population III remnant black holes grow to no more than about an order of magnitude larger than the initial seed mass at $z=0$.  Massive seed masses remain effectively unchanged.  This is an important result, independent of the formation scenario.  MBHs in dwarf galaxies indeed provide a clear indication of the initial seed properties.

This can be seen in Figure~\ref{fig:BHOF}, where we show the expected relationship between MBHs and their `dwarf' hosts. At $\sigma<20-40\kms$ there is no correlation between MBH masses and velocity dispersion. This is due to the very limited mass growth of MBHs hosted in low-mass satellites. These satellites have a very quiescent merger history (very few satellites experience any major merger at all at $z<6$), causing their central MBHs to starve and remain near the original seed mass, creating an asymptotic horizontal `plume' \citep[see also][]{VN09}. Additionally, a secondary role is played by mass stripping, that depends on both the mass of the host and the satellite, but also on the orbital parameters -- stripping being much more effective for radial orbits where satellites plunge deep into the potential well of the host, where densities are higher. A severely stripped system has a much larger MBH-to-host ratio. The final MBH-to-host relationship is therefore a combination of nature (formation mechanism) and nurture (MBH feeding and dynamical evolution of the host).

\section{Observational Prospects}

 \begin{figure}   
   \includegraphics[width=\columnwidth]{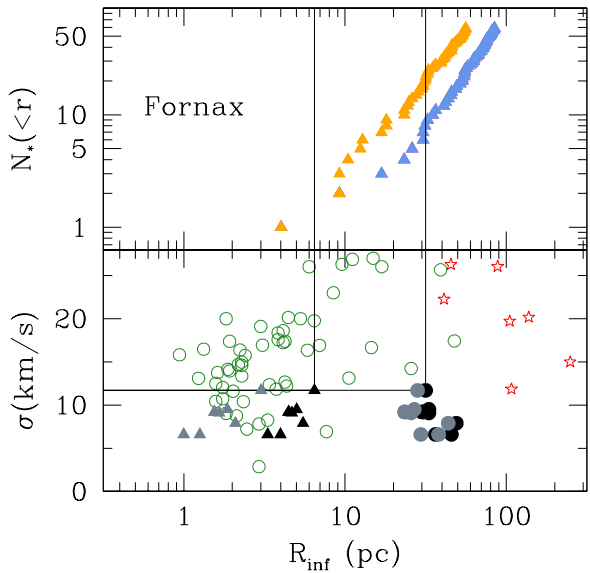} 
  \caption{Top panel: number of stars within a given projected radius in Fornax. Lower curve: all Fornax member stars for which velocities are currently available in the published kinematic samples of Walker et al. (2009a). Upper curve:  all Fornax target candidates, including unobserved stars, that are sufficiently bright ($V\la 21.5$) for velocity measurements with existing 6m -- 10m telescopes.   
      Bottom panel: relationship between velocity dispersion ($\sigma$) and radius of the sphere of influence of MBHs -- defined as the sphere that encompasses $2\times M_{BH}$ -- for ten halo realisations. Stars: `massive seeds'. Circles: `peak3.5'. 
      Black triangles: Milky Way satellites: we assume that the MBH sits on the $M_{BH}-\sigma$ relationship, and the density profile is cored. Grey triangles: as above, but for an NFW profile. 
      Black dots: we assume a fixed MBH mass, $10^5\msun$ and a cored profile. Grey dots: as above, but for an NFW profile.}
   \label{fig:rinf}
\end{figure}

\subsection{Dynamical measurements}
The presence of an MBH can be tested dynamically if the region where the gravitational potential of the black hole dominates the gravitational potential of the host can be resolved. This region is referred to as the sphere of influence of the black hole. We adopt here the definition of the sphere of influence as the region within which the enclosed mass in stars and dark matter equals twice the MBH mass. The radius of the sphere of influence is therefore defined as: $M(r<R_{\rm inf})=2\times M_{BH}$.

The lower panel in Figure~\ref{fig:rinf} plots stellar velocity dispersion against $R_{\rm inf}$, estimated for massive seed and Pop III cases.  Overplotted are values estimated for the eight `classical' dwarf spheroidal satellites  of the Milky Way, for which line-of-sight velocities have been measured for up to a few thousand stars per galaxy \citep{walker09a}.  For these objects we adopt the stellar velocity dispersion measurements of \citet{Walker2009}, and then adopt an MBH mass from the mass-velocity dispersion relationship \citep{Tremaineetal2002}.  In order to calculate the sphere of influence for the real dSphs, we consider the best-fitting cored and cusped (NFW) mass profiles with parameters listed in Table 3 of \cite{Walker2009}.  If we assume the observed dSphs have NFW dark matter profiles, then according to the M-sigma relation, their MBH masses correspond to spheres of influence of $1\la R_{\rm inf}\la 2$ pc.  If the dSph dark matter profiles are cored, then the spheres of influence are a few times larger, $3\la R_{\rm inf}\la 7$ pc.  

In order to evaluate prospects for detecting kinematic signatures from such MBHs in real dSphs, the upper panel of Figure \ref{fig:rinf} indicates the number of spectroscopic target stars within a given projected radius in Fornax, the most luminous dSph satellite of the Milky Way.  Curves indicate the cumulative surface brightness profiles of 1) all Fornax member stars for which velocities are currently available in the published kinematic samples of Walker et al. 2009a \citep[see also][]{Battaglia2006}, and 2) all Fornax target candidates, including unobserved stars, that are sufficiently bright ($V\la 21.5$) for velocity measurements with existing 6m -- 10m telescopes.  The latter profile represents the largest samples that are possible at present; unfortunately, these would include fewer than 5 stars within the spheres of influence estimated for the classical dSphs.  Thus even for the brightest nearby dSphs, the detection of any MBH must await the next generation of 20-30m telescopes, which may increase kinematic sample sizes by more than an order of magnitude.

Finally, we consider the spheres of influence due to MBH masses of $\sim 10^5\msun$, a mass of order of the upper limits derived for the `massive seed' scenario \citep[we recall here that the formalism proposed by \cite{LN2006} and adopted here gives the {\it total} mass available for forming an MBH seed. Most likely the final seed mass is smaller, see][]{BVR2006}.  For such masses, the expected spheres of influence reach $\la 50$ pc for the observed dsphs.  For Fornax, the expected value of $R_{\rm inf}\sim 30$ pc encloses 10 stars in the existing velocity sample, and 25 stars in the list of current target candidates.  If all these stars are observed, the resulting sample may help diagnose whether Fornax has an MBH of mass $\sim 10^5\msun$.

Figure~\ref{fig:rinf} suggests that MBHs generated as `massive seeds', having larger masses and a larger $R_{\rm inf}$, would be the most favourable scenario from an observational point of view. However,  their typical BHOF is lower, being always below 40 per cent and decreasing to less than a per cent for 'true' dwarf galaxy sizes \citep[$\sigma \sim 5$--$15$ km s$^{-1}$;][]{Walker2009}. From Figure~\ref{fig:BHOF}, the typical BHOF for a massive seed MBH in a dwarf galaxy like Fornax is $\sim$ 2 per cent, meaning that $\sim$ 50 Fornax-like dwarfs would need to be observed to have significant probability of finding an MBH. Population III remnants have instead a higher BHOF, but their masses have not grown much since formation, making their detection harder.

\subsection{Gravitational waves}
The detection of gravitational waves from a MBH in a dwarf galaxy undergoing a merger with another black hole is another possible probe.  The Einstein Telescope, a proposed third-generation ground-based gra\-vi\-ta\-tio\-nal-wave (GW) detector  will be able to detect GWs in a frequency range reaching down to $\sim 1$ Hz \citep{Freise:2009}. Sources with masses of hundreds or a few thousand solar masses will generate GWs in this frequency range, which is out of reach of {\it LISA} or the current ground-based detectors.  Since dwarf galaxies  have a very quiet merger history, we would not expect many MBH-MBH mergers involving dwarf galaxies at the present epoch. However, gravitational waves may also be generated in dwarf galaxies by mergers between the central MBH and stellar remnants (stellar mass black holes) in the centre of the dwarf.   It is estimated that the Einstein Telescope might detect as many as one thousand of these inspiral events in globular clusters, assuming a relatively high intermediate mass black hole occupation fraction in the clusters~\citep{Gair2009}. The same mechanism would operate in dwarfs, and the predicted rate of mergers from this channel scales with the core stellar density, $n$, as $n^{4/5}$~\citep{Gair2009}. The core stellar densities in nearby dwarf galaxies are typically very low, however, e.g., the estimate for Fornax is $\sim 10^{-1}$pc$^{-3}$~\citep{Mateo:1998} and for Sagittarius is $\sim 10^{-3}$pc$^{-3}$~\citep{Majewski:2005}. These densities are several orders of magnitude lower than the typical core densities in globular clusters, which are $\sim10^{5.5}$pc$^{-3}$. 

We estimate the number density of MBHs in dwarfs via recent theoretical works that study the population of dwarfs as satellites of the Milky Way \citep{Reed2005,Diemand2007,Springel2008}. These simulations suggest that the number of satellites per halo has the following form:
\beq
N(>v_{\rm sat})=N_*\left(\frac{v_{\rm sat}}{v_{\rm host}}\right) ^{\alpha},
\eeq
where $v_{\rm sat}$ and $v_{\rm host}$ are the maximum circular velocity of the satellite and the host halo, respectively. \cite{Springel2008} find $N_*=0.052$ and $\alpha=-3.15$.  If we extrapolate the $M_{\rm BH}-\sigma$ correlation to $10^2$--$10^3\,M_\odot$ BHs and assume an isothermal galaxy profile, then $v_{\rm sat}\sim 10$--$20$ km~s$^{-1}$. With this formalism we obtain the number of satellites in the interesting mass range per dark matter halo ($N_{\rm sat}$), where the mass of the halo is uniquely determined by its maximum circular velocity.  The number density of dark matter halos can be easily obtained by integrating the modified Press \& Schechter function \citep{Sheth1999} which provides the mass function of halos, $dn/dM_h$. Therefore we estimate a number density of satellites (per comoving cubic Mpc) as:
\beq
n_{\rm sat}=\int \frac{dn}{dM_h} N_{\rm sat}(M_h) dM_h\sim 1 \, {\rm Mpc}^{-3},
\eeq
adopting the normalization proposed by \cite{Springel2008}.  We now have to correct for the fact that not all dwarf galaxies are likely to host an MBH. Based on the models presented here, a fraction $\sim 0.01$--$0.1$ of dwarfs host an IMBH with mass $\sim 10^2$--$10^3\,M_\odot$.  Furthermore, we must account for the fact that many of these satellites do not form stars \citep[and references therein]{Bovill2009}. Based on \cite{Maccio2010}, we estimate that a fraction $f_*\sim 0.8$ of dwarfs with masses $>2\times 10^7 \msun$ formed stars and will contain stellar mass BHs that can merge with the central IMBH.  The final estimate for the number density of dwarfs hosting an MBH that could be Einstein Telescope sources is $\sim 0.008$--$0.08$ Mpc$^{-3}$.  This number density is approximately an order of magnitude smaller than the $\sim 0.3$ Mpc$^{-3}$ assumed for globular clusters in \cite{Gair2009}.  Combining the core stellar density and the number density scalings, we conclude that the rate of GW detections from this channel is likely to be $\ll 1$ per year. Therefore, although it is not inconceivable that third generation GW detectors will detect events originating in dwarf galaxies, any events will be serendipitous. 

An alternative way of detecting MBHs lurking in dwarf galaxies would be via their emission when accreting surrounding material, either from a companion star or gas available as recycled material via mass loss of evolved stars (Dotti et al. in preparation). 

\section{Conclusions}
We study the black hole population of dwarf galaxies in a Milky Way-type halo to find observational signatures of the MBH seeding and growth processes. Dwarf galaxies have a quiet merger history, leading to an MBH population at $z=0$ that has not evolved significantly from the seed population at high redshift. We consider two MBH formation mechanisms: a `massive seeds' model where seed form via gas-dynamical collapse and a Population III remnant model. We seed the progenitor halos of a Milky Way-type halo with MBH seeds and follow their evolution until $z=0$. Incomplete halo mergers at low redshifts produce a satellite population consistent with observated and simulated results. We study the MBH population of these satellites and find:
\begin{itemize}
\item Most dwarf galaxy MBHs have not grown significantly since seed formation at large redshift. Dwarf galaxies have low masses and tend to be baryon-poor, suppressing MBH growth through mergers and accretion. Measurements of MBH masses in dwarf galaxies would directly probe the seed masses. 
\item The `massive seeds' model produces rare but large seeds ($M \sim 10^5 M_\odot$). The resulting MBH population at $z=0$ occupies less than a few per cent of typical dwarf galaxies. These MBHs are sufficiently massive that they may be detectable from the central stellar kinematics in bright dwarfs such as Fornax.
\item The Pop III remnant model produces many small seeds ($M = 10^2 M_\odot$). This MBH population is more abudant at low redshifts than in the `massive seeds' model, but the low mass MBHs cannot currently be detected. The detection of this population must await the next generation of telescopes.
\end{itemize}

\section*{Acknowledgements}
SVW and MV thank Francesco Haardt for providing a code to calculate equilibrium temperatures based on different UV backgrounds.  They also thank Nick Gnedin and Andrea Maccio for helpful discussions.  MV thanks the Institute of Astronomy in Cambridge for providing a summer haven far from the maddening crowd.  MGW acknowledges support from the STFC-funded Galaxy Formation and Evolution programme at the Institute of Astronomy, University of Cambridge.  Support for this work was provided by the SAO Award TM9-0006X.

\end{document}